\newcommand\citep\cite
\documentclass{iau}
\usepackage{graphicx}
\usepackage{natbib}

\title[Pre-planetary nebulae ] 
{Pre-planetary nebulae: a context for principles, progress, and  questions on how binaries and magnetic fields produce jets}

\author[Eric G. Blackman]  
{Eric G. Blackman$^{1,2}$}
\affiliation{$^1$Department of Physics and Astronomy, University of Rochester \\ Rochester, NY, 14621, USA \\ email: {\tt eric.blackman@rochester.edu} \\[\affilskip]
$^2$Laboratory for Laser Energetics, University of Rochester\\ Rochester, NY, 14623, USA}

\pubyear{2022}
\volume{366}  
\pagerange{}
\jname{The Origin of Outflows in Evolved Stars}
\editors{L. Decin, A.A. Zÿlstra \& C. Gielen, eds.}
\begin{document}

\maketitle

\begin{abstract}
Astrophysical outflows treated initially as spherically symmetric often show evidence for asymmetry once seen at higher resolution.  The preponderance of aspherical  and multipolar planetary nebulae (PN) and pre-planetary nebulae (PPN) was evident after many observations from the Hubble Space Telescope.  Binary interactions have long been thought to be essential for shaping asymmetric  PN/PPN, but how?  PPN are the more kinematically demanding of the two, and warrant particular focus. I address how  progress from observation and theory suggests two broad classes of accretion driven PPN jets: one for wider binaries (PPN-W) where the companion is outside the outer radius of the giant and accretes via Roche lobe overflow, and  the other which occurs in the later stages of common envelope  evolution  (CEE) for close binaries (PPN-C).  The physics within these scenarios  connects to progress and open questions about the  role and origin of  magnetic fields in the engines and in  astrophysical jets  more generally. 

\keywords{(ISM:) planetary nebulae: general; ISM: jets and outflows; stars: magnetic fields;stars: winds, outflows; stars: AGB and post-AGB; (stars:) binaries (including multiple): close}
\end{abstract}

\firstsection 
\section{Introduction and overview}


Planetary nebulae (PN) are the penultimate evolutionary state of  stars below $\sim 8M_\odot$
\cite[e.g.][]{Balick+Frank02}
 for stars massive enough to evolve off the main sequence during the age of the Universe. They are characterized by an ionization nebula sourced  by  photons from the hot exposed white dwarf (WD) at their core.  Typically, the PN phase lasts of order $10^4$ yr after which the remnant WD is left. Planetary nebulae have size scales  $\sim 10^{17}$ cm and ages $\sim 10^4$ yr. 
Pre-Planetary  nebulae (PPN) are reflection nebulae typically an order of magnitude smaller and an order of magnitude younger. They are more powerful in mechanical luminosity than PN,  and are likely  the  earlier stage of a  PN before sufficient mass clears to expose the ionizing core.

Early models of PN/PPN were  based on  the spherical interacting stellar wind (ISM) paradigm  \citep{Kwok1978}
where a slow wind is followed by a  fast wind  and the interaction produces a shocked bubble. 
Evidence of  asymmetry and bipolarity emerged  later 
\citep{Feibelman1985,Gieseking+1985,Miranda+1990,Lopez+1993}. This fostered generalizations of ISW models (GISW)  with density asymmetries from equator to pole 
to explain the asphericity  \citep{Balick+1987,Soker+1989}. Basic ingredients of GISW models are  present in more modern scenarios and simulations  involving   binaries, magnetic fields, and jets   \citep{Soker+2000,Blackman+2001,Soker2002lobes, Balick+2020,Garcia-Segura+2021,Ondratschek+2021}. 

From the mid  1990s,  high resolution
HST observations revealed that asymmetry was the rule rather than the exception and the statistical categorization of morphologies became clearer. As a population, PN are overall 80\% aspherical,  up to 1/2 of the latter  exhibiting jets \citep{Balick+Frank02,DeMarco+2011}. 
The prevalence of asymmetric PPN among PPN is closer to 100\%,  with many showing  multipolar structure  \citep{Borkowski+1997,Sahai+Trauger1998,Sahai+2009}.

The influence of binaries on  shaping was proposed from the early days
\citep{Paczynski76,Soker+1989,Soker1994,Soker1997,Soker+1994,Reyes-Ruiz+1999,Blackman+2001} and is now considered  essential to explain the high fraction of asymmetric PN/PPN, particularly when angular momentum \citep{Soker2006,Nordhaus+07}  and outflow kinematics are considered. 
While PPN may represent a strongly
($\sim {\rm few} 100\ {\rm yr}$) 
collimated jet phase due to close binary interaction, binaries likely also explain the asphericity evolution in the giant stellar wind phases  that precede PPN \citep{Decin+2020}.  In general, binaries  induce  both equatorial and axial features.
Examples of equatorial features    include spiral arms and  crystalline dust \citep{Edgar+2008,Mauron+2006,Kim+2017} 
and axial features  include   winds and jets  perpendicular to the orbital plane \citep{Hillwig+2016}. 

Binaries also   supply   free energy for  flows which can amplify  magnetic fields \citep{Blackman+2001,Nordhaus+07,Garcia-Segura+2021,Ondratschek+2021}. This amplification may occur in an accretion disk onto a companion, a circumbinary disk, or a merger. The large-scale field that grows can in turn mediate  launching and collimation of jets. Magnetic collimation acts to concentrate the pressure of a flow on axis, but the magnetic structures must themselves be collimated by ambient inertial envelopes or tori, as unbounded magnetic structures are  unstable.
 

Observationally, $>200$ PN have binaries \citep{Boffin+2019,Jacoby+2021}
and an estimated $\sim 20\%$ of PN are preceded by close enough interaction to have incurred common envelope evolution (CEE) \citep{Miszalski+2009}. 
The number of observed PN/PPN with binaries is less than the 
$50\%$ fraction for all stars \citep{Miszalski+2009,Raghavan+2010},  but this is a lower limit on the prevalence of binary influence. In fact it is 
plausible that all aspherical PPN/PN are influenced by companions
\citep{Ciardullo1999,Bond2000,Moe+2006,Soker2006,Corradi+2012,Jones2020,Decin+2020}  including planets \citep{Soker1996planets,Reyes-Ruiz+1999,Nordhaus+Blackman06,DeMarco+2011,Decin+2020,Chamandy+2021},
particularly if we allow for the fact that the binary companion  may be destroyed by tidal disruption  during or after playing   a dynamically significant role.
 
\section{Kinematic demands on   PPN/PN}

PPN warrant special focus because they are kinematically
more demanding than PN, and therefore provide stronger limits
on source engines if indeed PPN are earlier stages of PN.
The difference in kinematic demands is evident in  comparing the approximate values for PPN and PN below.
PPN fast wind durations are $\Delta t=100-1000$ yr; outflow speeds $v\gtrsim 50$ km/s;
mass in nebula $M=0.5M_\odot$; mass-loss rate ${\dot M}_f=5\times 10^{-4}\ M_\odot$/yr; momentum injection rate
$\Pi\sim 5\times 10^{39}\ {\rm g\cdot cm/s}$; mechanical luminosity
$L_{f}\ge 8\times 10^{35}$ erg/s.
For the PPN slow winds,  the corresponding values are
$\Delta t=6\times 10^3$ yr, $v\sim 20$ km/s;
$M=0.5M_\odot$; ${\dot M}_f= 10^{-4}M_\odot$/yr; 
$\Pi\sim 2\times 10^{39}\ {\rm g\cdot cm/s}$; 
$L_{s}\sim 10^{34}$ erg/s.

For comparison, the PN fast wind properties are: $\Delta t=10^4$ yr, $v\sim 2000$ km/s; $M=10^{-4}M_\odot$; ${\dot M}_f=10^{-8}M_\odot$/yr;
$\Pi\sim 4\times 10^{37}\ {\rm g\cdot cm/s}$; 
$L_{f}\ge 1.3\times 10^{34}$ erg/s.  
For the PN slow wind $\Delta t=10^4$ yr, $v\sim 30$ km/s; $M=0.1M_\odot$; ${\dot M}_f= 10^{-5}M_\odot$/yr; 
$\Pi\sim 6\times 10^{38}\ {\rm g\cdot cm/s}$; 
$L_{s}\sim 3\times 10^{33}$ erg/s.

Most importantly, for PPN jets where sufficient data has been obtained, 
mostly all have momenta per unit time satisfying
$d\left(Mv_j\right)_{PPN}/dt>1000L/c>d\left(Mv_j\right)_{PN}/dt>L/c$
\citep{Bujarrabal+01,Sahai+2009,Sahai+2017}. 
Thus, not only do PPN have higher momenta injection rates than the PN, 
these  rates cannot be explained
by optically thin radiative driving of jet outflows \citep{Bujarrabal+01}
and thus need another source,  most likely involving 
accretion and close binary interaction  \citep{Blackman+2014}. 


\subsection{Momentum constraints and accretion modes }
\label{s2.1}
The stringent momentum requirements of  PPN outflows allow constraining the engine paradigm as follows \citep{Blackman+2014}.
The mechanical luminosity of all non-relativistic astrophysical jets obeys
\begin{equation}
L_{m}=\frac{1}{2}{\dot M}_{j,N} v^2_{j,N}\le  \frac{1}{2} \frac{G M_a\dot M_a}{r_{in}}=\frac{1}{2}{\dot M}_av_K^2\\
\label{2.1}
\end{equation}
where ${\dot M}_{j,N}$ is the "naked" jet mass ejection rate,
$v_{J,N} \equiv Q v_K(r_{in})$,  
 is the ``naked" jet launch speed, 
 $v_K=(GM_a/r_{in})^{1/2}$ is the Keplerian speed at the inner radius $r_{in}$ of the assumed accretor, $Q$ is a dimensionless number with a typical range  $5\gtrsim Q \gtrsim 1$ 
 in MHD jet models
 \cite[e.g.][]{blandford1982,Pelletier+1992}, 
 and ${\dot M}_a$ is the accretion rate.
Inequality (\ref{2.1}) and the definition of $Q$ imply 
\begin{equation}
 {\dot M}_a>Q^2 {\dot M}_{j,N}. 
 \label{2.2}
\end{equation}
As the  jet runs into surrounding material, 
momentum conservation implies ${\dot M}_{j,N}Qv_k={\dot M}_{j,obs}v_{j,obs}$, where ${\dot M}_{j,obs}$, and $v_{j,obs}$ are the observed jet mass ejection rate and speed that account for mass pileup. Solving for ${\dot M}_{j,N}$, and plugging into inequality  (\ref{2.2})  allows us to obtain a minimum for ${\dot M}_{j,N}$ and thus   ${\dot M}_a$ given by  
\begin{equation}
    {\dot M}_a>
    Q^2 \frac{{M}_{j,N}}{ t_a} s\simeq
    Q \frac {M_{j,obs}v_{j,obs}}{ v_K t_a},
\end{equation}
where $M_{j,obs},\ v_{j,obs}$ and $t_a$ are the observed mass, speed and lifetime measured from observations, and we have used ${\dot M}_{j,obs}=M_{j,obs}/t_a$.
Numerically we have  
\begin{equation}
{\dot M}_a\ge 1.4\times 10^{-4}\frac{M_\odot}{{\rm yr}}\left(\frac{Q}{3}\right) \left(\frac{M_a}{M_\odot}\right)^{-1/2}\left(\frac{r_{in}}{R_\odot}\right)^{1/2}\left(\frac{M_{j,obs}}{0.1M_\odot}\right)\left(\frac{v_{j,obs}}{100{\rm km/s}}\right)\left(\frac{t_{a}}{500{\rm yr}}\right)^{-1}.
\end{equation}

So which modes of accretion  satisfy this constraint?  The values for a number of accretion modes together with the requirements for specific PPN are shown in Figure \ref{figurelucchini}.
Bondi-Hoyle-Lyttleton (BHL) onto secondary does not work because
typical values would be 
  $\dot M_{BH}=1.15\times 10^{-6}{\dot M}/{\rm yr}$ \citep{Huarte-espinosa+13}, for a primary wind of mass-loss rate ${\dot M}_W=10^{-6}M_\odot/{\rm yr}$, wind speed $v_W=10$ km/s, primary mass $M_p=1.5M_\odot$ and secondary mass $M_s=M_\odot$.
Wind Roche lobe overflow (WRLOF)
\citep{Mohamed+Podsiadlowski12,Chen+17,Chen+18} 
also does not work for a typical
separation $a\simeq 20$AU,  dust acceleration radius
$R_d\simeq 6R_p\simeq 10{\rm AU}$, primary Roche lobe  radius $\sim 8.5{\rm AU}$, ${\dot M}_{WR}=2 \times 10^{-5}M_\odot/{\rm yr}$, and
 component masses $M_s=0.6M_\odot$ and $M_p=M_\odot$.

The semi-empirically determined  accretion rate for the Red Rectangle PPN 
is ${\dot M}_{RR} \ge  5 \times 10^{-5}M_\odot/{\rm yr}$, based on the luminosity of the far UV continuum of its HII region  \citep{Jura1997,Witt+2009}.  
This  is also too small for most PPN.

There are two classes of accretion modes that do work. First, Roche lobe overflow (RLOF) \citep{Meyer+1983,Ritter1988}   onto the companion from the primary envelope for a companion located outside the giant's envelope. 
For typical parameters, the RLOF rate is ${\dot M}_{RL}\simeq \rho_e R_e c_{s,e}^3/(GM_G)\gtrsim 10^{-4}M_\odot/{\rm yr}$
which gives a marginally sufficient lower limit, where $\rho_e,R_e,c_{s,e}$ are the density, radius and sound speed of the outer giant envelope, and $M_G$ is its mass.

A second viable  accretion mode  involves close binary interaction subsequent to when the secondary enters
CE.  This will only be successful after much of the envelope is unbound as I now explain. Accretion onto the core of the primary \citep{Soker+1994}
or  the secondary within  CEE
 can in principle be super-Eddington during the plunge phase of CE before the envelope is unbound, 
as estimates and simulations suggest  
${\dot M}_{CE}\ge  10^{-3}M_\odot/{\rm yr}$, for 
secondary and primary masses $M_s=0.6M_\odot$ and $M_p=M_\odot$, respectively
\citep{Ricker+Taam12,Chamandy+18}.
But such modes require a ``pressure valve” at the engine, otherwise accretion will be  halted. Were the accretor  a neutron star, neutrinos could supply this valve.  For main sequence or WD accretors 
the jet itself must supply the valve.  
Estimates of the ram pressure and simulations show that a jet from a main sequence or  WD companion during the plunge stage of  CE is likely choked within the bound envelope 
\citep{Lopez-camara+2021,Zou+2022}.
Accretion onto the primary core,
or accretion from a 
shredded low-mass companion 
\citep{Reyes-Ruiz+1999,Blackman+2001,Nordhaus+Blackman06} or circumbinary accretion  or merger are more effective \citep{Ricker+Taam12,Garcia-Segura+2021,Ciolfi2020,Ondratschek+2021}. These all take advantage of the deep gravitational potential well of an accretor and a substantial mass supply. The jet would be  visible in the later stages  when the CE envelope is unbound. The jet propagates along a reduced density  axial channel
\citep{Zou+2020,Garcia-Segura+2021,Ondratschek+2021}.

\subsection{Energy constraints and time sequence}
Energy constraints on PPN are also revealing. In cases measured, the  energy in PPN outflows typically exceeds the envelope binding energy of the AGB host stars  and exceeds the orbital energy from inspiral to observed radii from CE  \citep{Huggins2012,Olofsson+2015}.
This also points to the need for accretion or a merger to  tap into the deeper gravitational potential wells, or perhaps even nuclear energy.

While PN, unlike PPN, do not strongly constrain the energy or momentum, both PPN and PN do  mutually provide  a time sequence constraint on the jet and equatorial torus. In different sources, PPN/PN jets are observed to occur both before  and after the equatorial dust tori form. In one sample, \cite{Huggins2007} found that  PPN jets follow tori by $\sim 250$ yr on average. That  is consistent with equatorial ejecta helping to facilitate collimation,  independent of whatever role magnetic fields might play. On the other hand,  \cite{Tocknell+2014} studied the kinematics of four post-common envelope PN. Three have jets that preceded CE ejection and one has 2 pairs of jets that follow the torus.  Although CE is one natural way to get an equatorial torus, a torus may also form from an earlier  RLOF  phase  \citep{Macleod+18b} as mass leaves through the L2 point and enters bound orbits.

\subsection{PN are plausibly later stages of accretion driven PPN} 

If an accretion-like  process onto the core of a companion of mass 
$M_*$ powers PPN jets, then a connection between PPN and PN is kinematically consistent: the jet mechanical luminosity is
\begin{equation}
L_{m} \simeq \frac{GM_{*,\odot}{\dot M}_a\epsilon}{2R_i}=4.5\times 10^{36} 
\epsilon_{-1}
\left(\frac{M_{*,\odot}{\dot M}_{a,-4}}{R_{i,10}}\right),
\end{equation}
where $\epsilon_{-1}$ is a dimensionless efficiency from accretion to jet power in units of $0.1$;  $M_*,\odot$  is the accretor mass scaled in solar masses; ${\dot M}_{a,-4}$ is the mass accretion rate scaled in units of $10^{-4}\ {\rm M_\odot/ yr}$;
and $R_{i,10}$ is the inner disk radius in units of $10^{10}\ {\rm cm}$.
The naked jet speed is
\begin{equation}
  v_{j,N}\sim Q v_{K}\simeq 1600 Q \left(\frac{M_{*,\odot}}{R_{*,10}}\right)^{1/2}{\rm km/s},
\end{equation}
so that the predicted observed PPN jet speed after mass pile-up from momentum conservation when the ejecta are optically thick, is given by
\begin{equation}
   v_{j,obs} \simeq \frac{M_f v_f}{f_\Omega M_{env}+M_f}\sim 80{\rm km/s},
\end{equation}
where $M_f=M_{j,N}$ and $v_f=v_{j,N}$.
But once the outflow transitions to the PN stage the optical depth $\tau_d$ to dust scattering drops below unity, as estimated by 
\begin{equation}
\tau_d=2.5\times 10^{-3}   \left(\frac{n_d}{2.5\times 10^{-13}{\rm cm^{-3}}}\right)\left(\frac{\sigma_d}{10^{-8}{\rm cm^2}}\right)\left(\frac{R}{10^{18}{\rm cm}}\right),
\end{equation}
where the dust number density $n_d$ and scattering cross section $\sigma_d$ are scaled to typical PN values.
This is a  simple explanation for the trend that PN have less power but  faster winds. The naked  jet,  and thus the naked jet speed, is more exposed in PN as the optical depth decreases.
More detailed transitions from PPN to PN speeds and powers can be predicted  from engine models
as a function of age and compared to individual sources or statistical observations. 

\section{Binary interactions: from weak to strong for low mass giants and low mass companions}

Since the mechanisms  of accretion that work to power PPN require binary interactions with orbital radii at least small enough for the primary to overflow its Roche lobe, the question of how sufficient numbers of binaries   get close enough to produce the required number of PPN/PN
arises.    Observations suggest that at least 
 20\% of PN have to have incurred CE
\citep{Miszalski+2009}.  But  since  only 2.5\% of PPN/PN should incur CE if tides alone are responsible for orbital decay to the RLOF phase \citep{Madappatt+2016},  something else to tighten the orbits is needed.

For wide separations, analytic estimates of BHL accretion, which are too low to power PPN,  match simulations  and  an accretion disk forms around the primary primarily from infall toward the retarded position of the secondary   \citep{Huarte-espinosa+13,Blackman+2013}. 
But for this mode of accretion, such a high fraction of the mass lost from a typical AGB wind 
from which the BHL accretion draws, leaves  without interacting much with the secondary. Therefore the orbit tends to increase \citep{Chen+18,Decin+2020}.  If however, for  
somewhat tighter orbits,  
WRLOF occurs \citep{Mohamed+Podsiadlowski12},  
wind accelerated  orbital decay \citep{Chen+17,Chen+18} is possible and can greatly increase the number of systems that ultimately incur close enough interactions to produce PPN.  More work is needed to make exact predictions.

\begin{figure}[b]
\vspace*{-5.0 cm}
\begin{center}
 \includegraphics[width=5in]{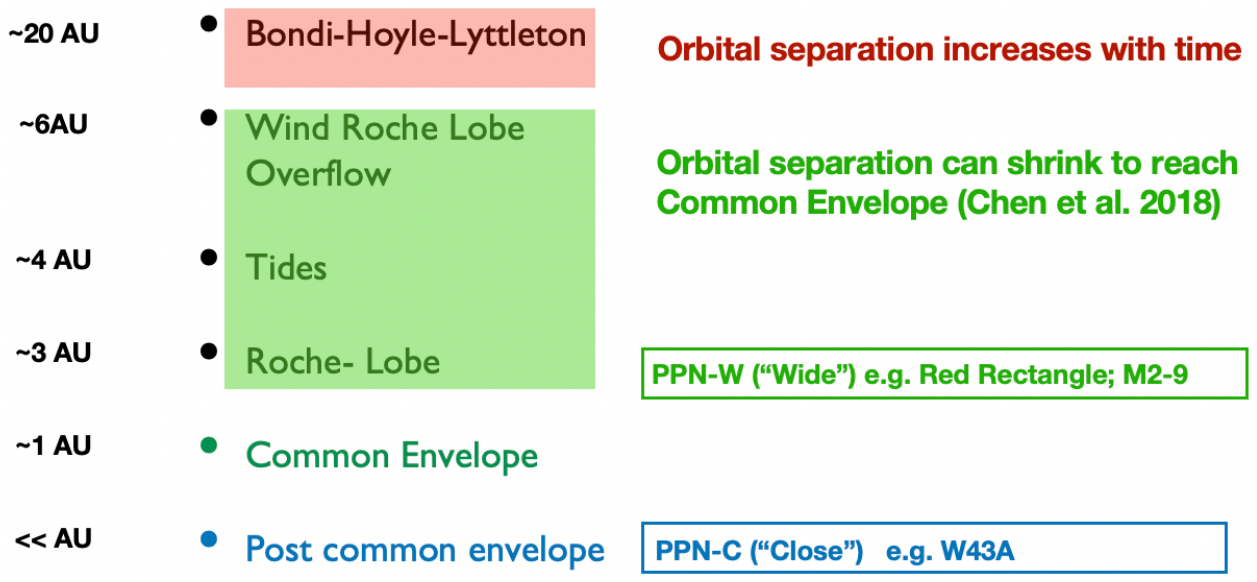} 
 \vspace*{-5.0 cm}
 \caption{With decreasing orbital separation from top to bottom, the figure schematically indicates that separations that are too wide initially will widen further with BHL accretion. The red shading indicates no PPN will form from these systems.
In the green region, indicated by the dominant mechanisms listed, the orbital separation can shrink  \citep{Chen+18}.
PPN can be produced at two key stages: in the RL regime indicated for this plot around 1-3 au (corresponding to PPN-W)  and after much of the envelope is unbound in the post-CE stage (corresponding to PPN-C). Suggested examples of these two cases are shown in Figure \ref{figurem29w43a}.}
\label{figurechen}
\end{center}
\end{figure}

Figure \ref{figurechen} shows the different consequences of initial  binary separation, and  where a PPN can be powered.

\begin{figure}[b]
\vspace*{-.65 cm}
\begin{center}
 \includegraphics[width=4.1in, angle = -90 ]{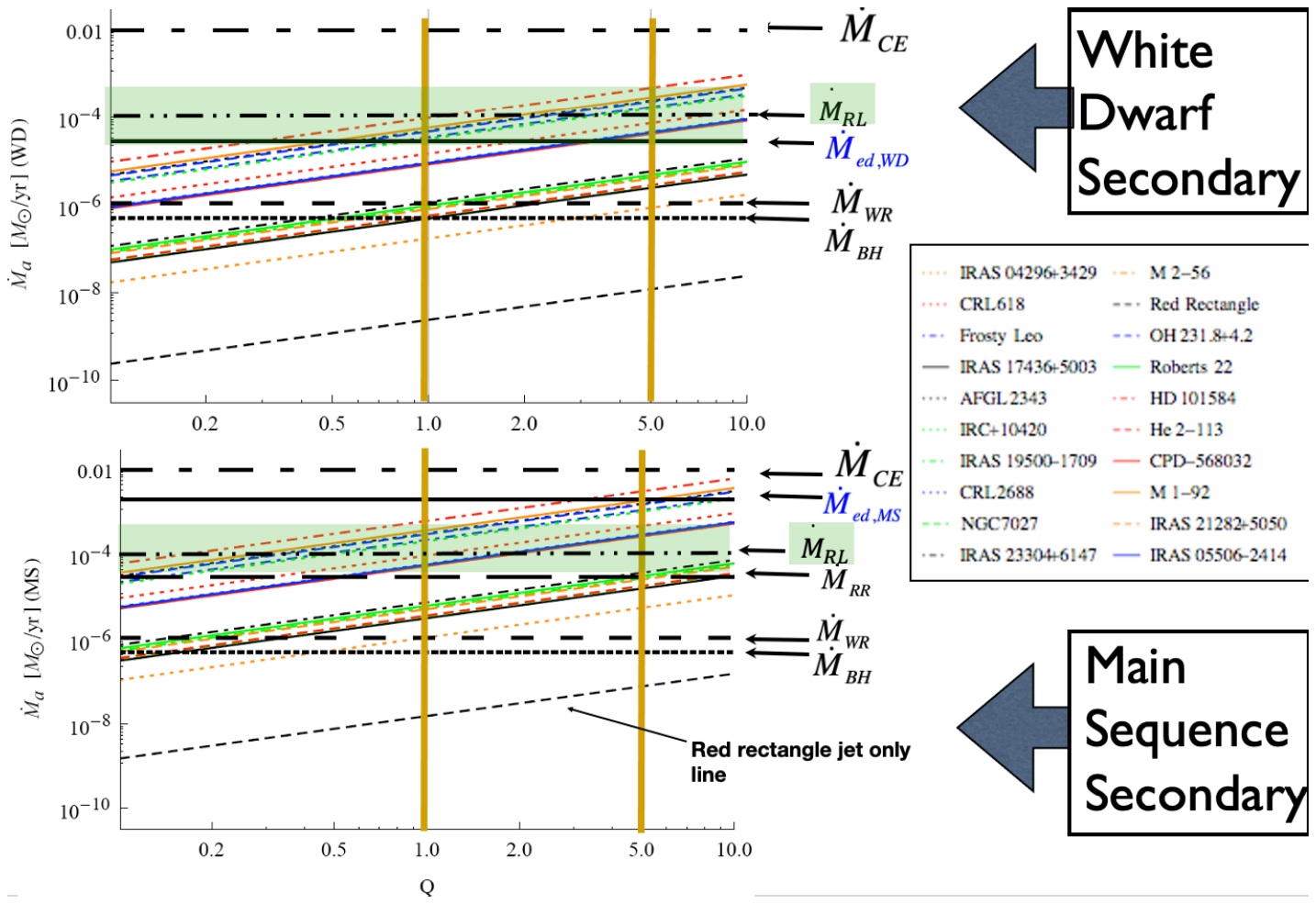}
 \vspace*{-.8 cm}
 \caption{Adapted from \cite{Blackman+2014}: Horizontal lines in each panel show examples of theoretically estimated accretion rates  (BHL $M_{BH}$; Wind Roche lobe overflow $M_{WR}$; Red Rectangle based on HII region $M_{RR}$; Roche lobe overflow $M_{RL}$; Eddington $M_{ed}$; Accretion deep in common envelope $M_{CE}$) and the diagonal lines correspond to momentum requirements inferred from observations for each of the objects in the table as $Q$, the  ratio of naked jet speed to Keplerian speed at the disk inner radius, is varied. The  vertical gold lines  bound the typical range of $Q$ from MHD jet models. The accretion mode of a given horizontal line is  sufficient to power jets only where the points on a given horizontal  line lie above those on a given diagonal line. The figure shows that RL and higher accretion rates are sufficient to power PPN but lower accretion rates are not generally sufficient.}
 \label{figurelucchini}
\end{center}
\end{figure}

\subsection{Distinction between "PPN-W" and "PPN-C" and Example Cases}

Two classes of mechanisms  work best to power PPN feature in the above discussion.   In the symposium talk, I distinguished them as ``preceding" and ``succeeding"  CE.  In the discussion with R. Sahai, J. Kluska, and N. Soker afterward, it was suggested to me that a better distinction might be AGB and post-AGB since there are  PPN objects   which have 100 to 1000 day orbital periods \citep{Bollen+2021}  and will never incur CE since their envelopes are mostly gone.  However, what I wish to convey in a classification scheme  is  distinguishing PPN mechanisms by their binary separation.  This   distinction can apply to systems with either AGB or RGB primaries. 

Taking all of this into account,  I label the two classes as PPN-W and PPN-C, where the W and C stand for ``wide" and ``close" and correspond, respectively,  to  orbital separations larger and smaller than the original giant envelope.  Thus PPN-W would include RLOF accretion PPN, and PPN-C refers to close binary mechanisms. Inasmuch as the PPN in the aforementioned \citep{Bollen+2021} objects depend on the  interaction of the observed binaries, these objects would be classified as PPN-W.  I now discuss some specific example objects, and subsequently review some simulations in this context.  

\begin{figure}[b]
\vspace*{-3.0 cm}
\begin{center}
 \includegraphics[width=5.5in, angle =0 ]{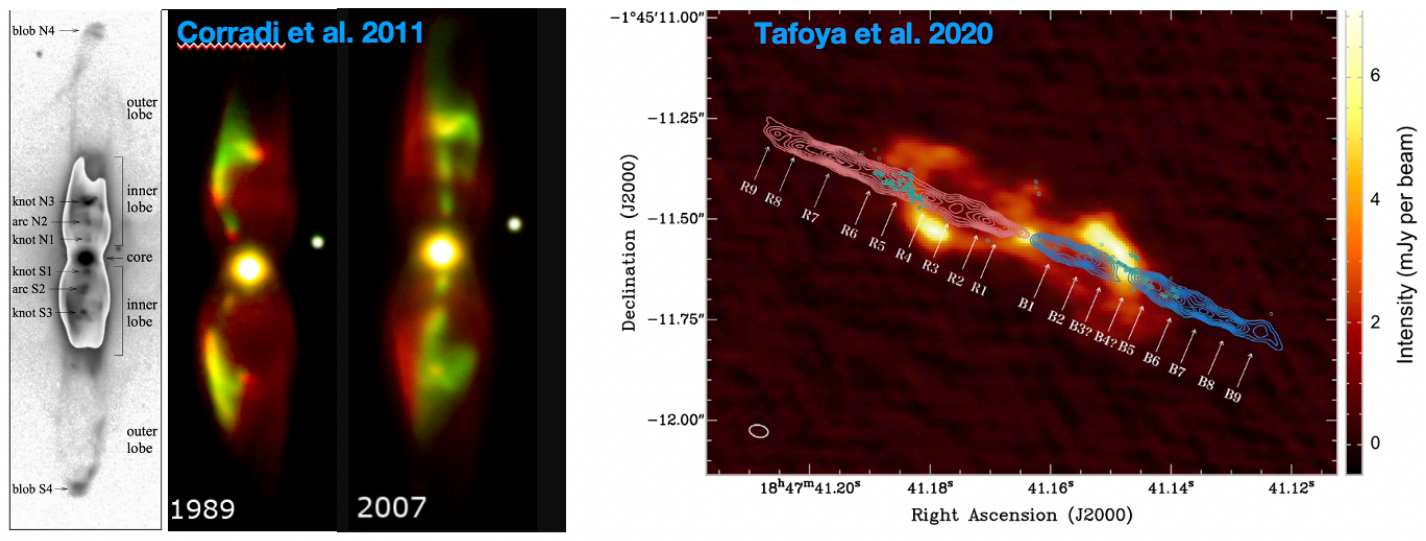}
\vspace*{-3.0 cm}
 \caption{Left panel shows images of M2-9   from \cite{Corradi+11} and right panel shows W43A  from   \cite{Tafoya+2020}.  M2-9 is  likely a PPN-W, as the jet is produced from  an accreting companion just outside the orbital radius of the giant, and consistent with RLOF accretion. Mirror symmetry is seen as the jet orbits and illuminates the surrounding bubble. In contrast, W43A is more likely a PPN-C, having a  tightly collimated straight jet, plausibly produced deep within an CE, after a companion plunged close to the primary core.}
 \label{figurem29w43a}
\end{center}
\end{figure}

Figure \ref{figurechen} shows schematically the binary separation that distinguishes these two PPN classes,  and  Figure \ref{figurem29w43a} shows two examples. Other examples  are discussed below. 


\subsection{Example PPN with the PPN-W and PPN-C distinction in mind}

\textit{HD 44179: Red Rectangle:}
The Red Rectangle PPN is best modeled as a main sequence secondary,
accreting from the primary giant in an elliptical orbit, and moving in and out of the Roche Lobe of the primary \citep{Witt+2009} and is  an example of a PPN-W. The outflow  emanates  within the  central cavity of a cicumbinary torus of thickness 90 au and cavity diameter of 30 au,  consistent with the formation of a torus from RLOF
 \citep{Macleod+18b}. 
 The HST composite from \cite{Cohen+2004} shows a  bipolar axis length of $\sim 1.5\times 10^4$ au. The jet produces a blue shift in H$\alpha$ emission, modulating the primary’s envelope emission.
 The  accretion rate as constrained by the  luminosity needed to source far-UV continuum for its HII region  \citep{Jura1997} assuming a distance of 710 pc,  implies a  maximum accretion disk temperature of 17,000K  and a minimum accretion rate  of ${\dot M}_a \ge  5 \times 10^{-5} M_\odot/$yr. This is is plotted in Figure \ref{figurelucchini} and is much larger than the constraint purely from the jet momentum calculation above, but can be accommodated by RLOF.



\textit{M2-9:}
M2-9 is  a PPN-W with a jet from a companion orbiting in an 88-120 yr period,  and injecting the flow  into a hot bubble cavity. 
Early binary scenarios \citep{Soker+2001}  
are now updated as \cite{Corradi+11} favored a fast jet induced illumination of the inner cavity   \citep{Doyle+2000}  rather than a photon source, based on delay time between knots.
The jet produces  mirror symmetry. There has been some ambiguity in interpreting this as potentially a  symbiotic (RGB + WD companion instead of AGB+WD/MS) but the distinction is  not  important from a basic theoretical jet mechanism perspective since accretion from the RGB envelope onto the companion vs. accretion from the AGB envelope onto the companion both represent "W" nebulae in the classification  above. 
\cite{Lykou+2011} also identify a  “disk/torus" from 15-900 au which
may  play a role in  collimating the flow.

\textit{W43A:}
W43A is one of 15 ``water fountain" sources exhibiting 
water maser emission
\citep{Diamond+1988,Imai+2002,Imai+2005,Vlemmings+2006,Amiri+2010,Chong+2015}.
\cite{Tafoya+2020} interpreted W43A ALMA data in CO to reveal knots  separated by a  few years,  a jet launched at 175 km/s, decelerating to 130 km/s,  and collimated from 90 au to 1600 au.
There is no   binary detected, but  \cite{Tafoya+2020} suggest that the knots could indicate the influence of a binary period e.g. an eccentric orbit. That would make this a PPN-W.  However, the  tight collimation suggests that  the outflow is produced from  a tighter binary engine, suggesting that it may  be a PPN-C outflow. The knots would then indicate a secular time scale (perhaps due to unsteady viscous accretion) compared to the much shorter orbital time scale at the base of the jet. 

\cite{Garcia-Segura+2021}  also argued that W43A is a post-CE  object, also making it a PPN-C.   In fact \cite{Khouri+2021}  observed that W43A and  the other 14 known water fountain sources all likely incurred CEE, which implies that they are all PPN-C.

Magnetic fields measured in W43A  are $85$ mG at $\sim 500$ AU and are strong enough to collimate the measured outflow on those scales \citep{Vlemmings+2006,Amiri+2010}.
These fields are quite far from the  likely jet origin, but if scaled even linearly down to $10^{11}$ cm could provide the $>$ kG fields needed to be dynamically significant at the engine.  Whether the source of the fields at 500 au is separate from, or an extension of, the fields generated in the jet engine  is not yet clear.

\textit{IRAS16342-3814:}
This is another water maser fountain source
with a  collimated molecular jet and dust emission \citep{Murakawa+2012,Sahai+2017}.  As observed by \cite{Sahai+2017}, its  high-speed   jet exhibits 5 knots/blobs in each lobe, and gas of density $\sim 10^6/{\rm cm^3}$ is expanding in a 1300 au torus.
 There has been a rapid increase of mass-loss rate
  to  $>3.5\times 10^{-4} M_\odot/{\rm yr}$  in the past $\sim 455$ years which suggests CEE.

\cite{Sahai+2017} also  constrain   
the circumstellar component ages for the 
AGB circumstellar envelope ($\sim 455$ yr); extended high velocity outflow (EHVO) ($130\ {\rm to}\ 305$ yr with speed $\sim 500$ km/s);  dust torus 
$(160\ {\rm yr}$);  high velocity outflow (HVO) $(\sim 110\ {\rm yr},$ with speed $\sim 250\ {\rm km/s})$. 
These indicate that the torus emerges several hundred years after the rapid AGB mass-loss increase, and the HVO appears very soon after torus formation. This is consistent with the time sequence in the \cite{Huggins2012} sample mentioned earlier.  

Although data from this object are not used in Figure \ref{figurelucchini}, the inferred kinematics  also require  accretion rates as high as RLOF from the primary,  or accretion operating  within/after CEE.
The high collimation, the absence of a detectable binary and the presence of a substantially increased AGB envelope all point to this source being classified as a PPN-C.

\textit{Calabash (Rotten Egg) Nebula OH232.84+4.22:}

This object has a binary engine consisting of a   
Mira (AGB) variable  \citep{Cohen+1981,Kastner+1998}, and an A0 main sequence companion in a likely 
$\ge 50$ au orbit \citep{SanchezContreras+2004}.
The orbit is too wide to provide the needed accretion rate $(\sim 10^{-4}M_\odot/{\rm yr})$ to power the $\sim 0.2$ pc bipolar nebular lobes and explain the rate at which  $\sim 1M_\odot$ of circumstellar molecular gas from previous mass loss arose. 
\cite{SanchezContreras+2004} speculate that an FU Orionis type outburst from accretion onto the companion  might account for this. The source would then be a PPN-W. 

However,  another possibility is that there was a previous binary inspiral via CE which ejected envelope material and powered the PPN as a PPN-C.
The outflow could then be sourced by  circumbinary accretion or the release of free energy due to pre-merger core activity, with energy released as heating or mediated by self-generated magnetic fields along
the lines of \cite{Ondratschek+2021} discussed  below. 
In this case, the presently observed wide binary would have little to do with what actually produces the collimated outflow. The Calibash nebula also has a rather  collimated spine with little wobble.  \cite{Sabin+2015}, using CARMA,  
found polarization that indicates  a  mostly toroidal ordered magnetic field perpendicular to the outflow. This is  consistent with the orientation expected if some magnetic collimation were at work.  


\cite{Dodson+2018} found that H$_2$O maser emission aligned along bipolar lobes is  perpendicular to an SiO maser disk  and is $\sim40$ yr old, confirming that mass loss is ongoing in the jet  and that the history of mass loss is  unsteady.


\section{Common Envelope Evolution}

CEE begins when the giant envelope engulfs the orbit of the secondary and the latter plunges in. The envelope could directly engulf the companion upon expansion to a giant phase for  initially small enough orbital radii, but for most systems that incur CE, 
the process  is likely preceded by a slow inward orbital migration that proceeds from  wind induced orbital decay, tides, and RLOF. As discussed earlier, although the  binaries that undergo BHL accretion expand  because so little of the mass lost from the primary interacts with the secondary, closer  binaries that incur stronger interaction via WRLOF can tighten. This tightening is further exacerbated by tides. 
CEE is ultimately important both for determining the properties and structure of mass loss as well as the orbital evolution of stellar systems that may include planets.
The resulting effect on binary evolution and the efficacy with which angular momentum is removed from the system influences compact object merger rates and
basic phenomenological properties of both high and low mass stellar  systems  \citep{Paczynski76,Iben+Livio93,Taam+Sandquist00,Ivanova+11,Ivanova+13b,Ivanova+13a,Postnov+2014}.  

CEE is challenging to model accurately because of the wide range of temporal, spatial, and density scales from the $1$ au envelope radius to the core dynamics $\le 10^{10}$ cm, let alone the nebular outflow extending out to 
$\ge 0.1$ pc if one is to follow the full influence of CEE  from giant star to PN.  CE simulation efforts have, however, been progressing with substantial progress \citep{Ricker+Taam12,Ivanova+13b,Ivanova+13a,Ivanova+15,Ivanova+Nandez16,Ivanova18,Ohlmann+16a,Ohlmann+16b,Ohlmann+17,Macleod+Ramirez-ruiz15b,Macleod+Ramirez-ruiz15a,Macleod+17,Macleod+18b,Macleod+18a,Staff+16a,Staff+16b,Iaconi+17,Iaconi+18,Iaconi+19,Iaconi+20,Chamandy+18,Chamandy+19b,Chamandy+19a,Chamandy+2020,Ondratschek+2021}.  Most simulations are run for $\sim100$ days
with the long end being $\sim4500$ days \citep{Ondratschek+2021}. Even that however, is short compared to the $100$ to $1000$ yr lifetimes of PPN, and the $\sim 10^4$ yr lifetimes of PN.

\begin{figure}[b]

 \vspace*{-3.0 cm}
\begin{center}
 \hspace*{-1 cm}
 \includegraphics[width=4.6in, angle =-90 ]{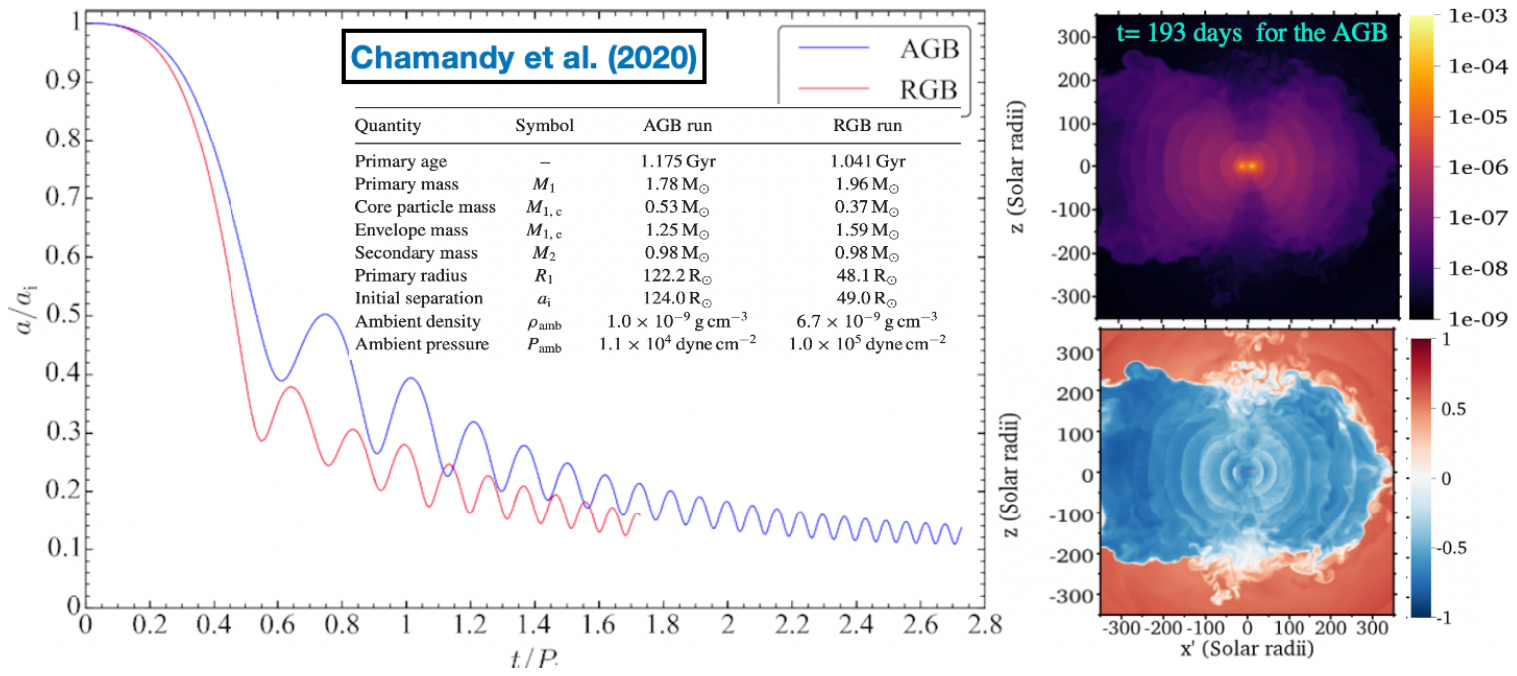}
\vspace*{-3.0 cm}
 \caption{taken from \cite{Chamandy+2020}. 
Left panel: inter-particle separation   for  AGB and RGB runs with the  rapid plunge
phase, followed by slow inspiral.
Time is normalized by orbital period:  96.5 days for the AGB curve and 23.2 days for the RGB. 
Separation is normalized by the initial orbital separation: $124\  {\rm R_\odot}$ for AGB and  $49 \ {\rm R_\odot}$ for the RGB. 
The AGB run is shown  to $\sim260$ days.   The rate of unbinding at the end is  steady at   $0.17 M_\odot/{\rm yr}$, and would unbind the full envelope within $7$ yr if extrapolated. Right panel: vertical slices through the orbital plane of the AGB at $t=193$ days for the AGB run. The top  slice is the gas density in ${\rm g/cm^3}$ and the bottom slice indicates bound (blue) and unbound (red) gas at this time quantified by a dimensionless measure ranging from 1 (strongly unbound) and -1 (strongly bound).}
 \label{figurechamandy}
\end{center}
\end{figure}

Important open questions include: how efficiently can a companion  of given mass  unbind the envelope upon inspiral? Does unbinding require recombination energy \citep{Soker04,Ivanova+Nandez16,Glanz+Perets18}?   What is the effect of convection?
Convection might reduce the efficiency with which recombination might supply energy \citep{Wilson+Nordhaus18} but might also redistribute energy more efficiently causing more mass to be unbound  
\citep{Chamandy+19a}.  
For simulations without convection and without recombination however,  CE may unbind the AGB  envelope  if the results from runs  of 260 days for a $1.8M_
\odot$ primary and $1M_\odot$ secondary  were extrapolated  to
$\sim 7$ years \citep{Chamandy+2020}, 
as shown in Figure \ref{figurechamandy}.

\cite{Armitage+Livio00} and \cite{Chevalier2012} examined the ejection of  CE by jets launched from a neutron star that inspirals inside the giant envelope. Others have considered different companions \citep{Soker04,Papish+15,Soker15,Soker16,Morenomendez+17,Shiber+17,Shiber+Soker18,Shiber+19,Lopez-camara+19,Lopez-camara+2021}.
As discussed earlier, 
accretion onto a plunging main sequence or WD companion before the envelope is unbound, requires a  pressure release valve in order to prevent  thermal pressure from building up and abating the accretion
\citep{Chamandy+18}.  
Jets from MS and WD companions are largely choked by the bound envelope after plunge-in for reasonable jet powers \citep{Lopez-camara+2021}. This can  be estimated by comparing the ram pressure of the jet with the thermal pressure of the envelope.  The total energy injected by the jet $L_{m}t_j$ where $t_j$ is the jet lifetime, can also be compared with the binding energy
to identify a minimum  time scale over which this outflow could unbind the envelope, compared to other unbinding  mechanisms. The efficiency with which the jet energy unbinds mass can also be studied with simulations by tracing the bound and unbound mass. Although the jet may be unimportant in a short simulation,  over longer times its effect may be more significant \citep{Zou+2022}.

Even if the jet starts  during RLOF before the secondary plunges, it will be connected by  an accretion stream to the envelope which will facilitate an already rapid drag-in to inspiral. Thus  it seems unlikely that  a jet would produce  much unbound mass before plunge in. For a limited set of binary parameters which  includes the case of a $1$ au primary of mass $18\ M_\odot$ and a $5.4\ M_\odot$ secondary with no jet, \cite{Macleod+18a} found that RLOF  can last decades but that once the secondary enters the envelope the plunge is rapid.  There have not yet been  numerical studies that include accretion from the envelope and include a jet  such that the envelope mass, accretion stream, and orbital inspiral are all self-consistently included,  but the lessons learned so far do not suggest that the jet could prevent inspiral. 



For CE,  the list of topics with opportunity for  more work includes:  (i) the inclusion  of convection; (ii) more complete treatments of the  equation of state and better approximations to inclusion of recombination energy;   (iii) more detailed radiative transfer; (iv) studying the sensitivity to  initial conditions and initial binary separation;  (v) increasing the duration of simulations;  (vi) limiting the effects of softening of potential wells by smoothing and limited numerical resolution;  (viii) ensuring conservation of energy and angular momentum in long term simulations. 

\section{Role of Magnetic Fields in Driving and Shaping}

Because collimated jets from MS and WD companions are substantially choked during CEE plunge-in as described above,  jets that produce PPN would be  most likely visible either in the  (i) RLOF phase before plunge producing a PPN-W or (ii) after substantial CE ejection has occurred producing a PPN-C.

So what about the collimation in each of these cases? Magnetic fields are likely important for jet launch and tight collimation  as  toroidal hoop stress can act to concentrate the pressure of the flow  to an axial spine.  But all magnetically collimated flows still require ambient pressure for stability. In fact there is no astrophysical context with a collimated jet for which an ambient  wind or thermal pressure surrounding the jet is ruled out.  In the present context,  CE provides tori  for inertial collimation both in the RLOF phase \citep{Macleod+18b,Macleod+18a} and in the post-CE ejection phase. We  discuss the latter  here.

\cite{Garcia-Segura+2018}, ran 2-D simulations for $10,000$ yr,  starting  at 1 au with  output from CEE simulation of \cite{Ricker+Taam12}  at a time  of 47 days as the initial conditions.
\cite{Zou+2020} ran 3-D simulations for 10,000 days, starting at  100 au with output from the  CEE simulation of \cite{Reichardt+19}.
The results from both of these simulations exemplify the basic principle  that an uncollimated hydrodynamic wind injected within the output of a CE simulation can be collimated  by the ejecta. However the outflow of  \cite{Garcia-Segura+2018} does not remain collimated and instead transitions to a wide barrel/elliptical structure in the absence of magnetic fields. 

Importantly, magnetic fields are  a ``drive belt”  not a ``motor” and require a source of  free energy (convection, orbital, accretion)   supplied by binary interactions. Magnetic fields and binaries are therefore not  competing mechanisms,  but operate in symbiosis. The binary supplies the free energy and sets up the environment within  which the magnetic field is amplified and functional.
Analytic estimates for dynamically important disk engine field strengths in the PPN context give values  $\ge 10$ kG \citep{Blackman+2001}.

Most astrophysical MHD jet simulations  have  separated the detailed origin of  the magnetic fields from the outflow dynamics. For example, the surface of the anchoring rotator has often been treated as boundary conditions with an imposed field.   But  progress is emerging across astrophysical disciplines in unifying field origin and jet formation self-consistently \citep{Kathirgamaraju+2019,Ruiz+2021,Ondratschek+2021}.
The range of approaches to the problem in the PPN/PN context has included  analytic calculations \citep{Pascoli1997,Blackman+2001,Tout+2003,Nordhaus+Blackman06,Nordhaus+07} 
focused on dynamo and/or  power generation  and three types of numerical
approaches discussed in the subsections below.

\subsection{Shaping from imposed  dynamically important magnetic fields}  

Shaping of  flows with an imposed magnetic field 
has been a long standing approach
to model some aspects of outflow shaping \citep{Chevalier+1994,Garcia-Segura1997,Garcia-Segura+1999,Garcia-Segura+2005,Balick+2020}.
\cite{Balick+2020}, for example, imposed a  flow of
 400 km/s with opening angle 40 deg  injected normal to a sphere of radius 1000 au. A toroidal field was imposed with initial values varying between  $0.003{\rm G} \le B\le  0.3 {\rm G}$. 
The framework was tuned to produce resultant angular distributions of speeds and outflow  geometry  consistent with e.g. OH231.8+04.2 \citep{Alcolea+2001,Sanchez-Contreras+2018}, 
 CRL618 \citep{Balick+2013,Riera+2014}, Hen2-104 \citep{Corradi+2001}, and MyCn18 \citep{Bryce+1997,O'Connor+2000}. 

Although such a method does not self-consistently generate jets because   a strong  magnetic field is imposed as an initial condition,  one can still use the results to match observations and then  infer what  the  field strength and geometry should be to inform further observations and theory.  
The method  has not been used to study the possible differences between PPN-C and PPN-W which would would be valuable.  

The next subsections focus on more steps toward self-consistent generation of collimated outflows, specifically in the PPN-C context.

\subsection{Multi-stage 2-D MHD simulations}  

A second approach is to use   output from 3-D CE simulations to set initial conditions for 2-D simulations, with only an  imposed weak seed field. The weak field may grow dynamically and produce  self-generated MHD outflows. 
Using this approach, \cite{Garcia-Segura+2020,Garcia-Segura+2021} 
started with the conditions of the \cite{Ricker+Taam12} CE after 56.7 days at which point 25\% of the envelope is unbound and mass is being lost at the high rate of 2$M_\odot$/yr.
A weak  seed magnetic field with toroidal and poloidal components was imposed on the scale of 1 au. The field grows, causes angular momentum loss, disk collapse, and a   magneto-centrifugally launched  wind. 
After 120 days, they took output for a second  2-D simulation, this time using an expanding grid envelope. 
They evolve the result for $> 1000$ years.
They find that the resultant CE outflow forms from a  circumbinary disk and a  very tightly collimated outflow whose
properties   
  plausibly resemble   W43A  and the Calabash nebula.
  This computational model produces a PPN-C in the aforementioned classification scheme.   The jet is produced after much of the envelope is ejected and the binary orbit is 7 times smaller than 
  the RGB envelope at the time of the conditions used from  \cite{Ricker+Taam12}.

This approach is a step toward more self-consistency in that the imposed magnetic field is weak and the outflow is self-generated. There are certainly limitations  to the fidelity of 2-D and the expanding grid, but the 2-D simulations can be run for $>1000$yr which is orders of magnitude  longer than what can be expected for  3-D  simulations.  There are always trade-offs, and precise realism is not always necessary to gain some insight. 

\subsection{Simulations with ``organic” 3-D magnetic field amplification and jet formation }  

Complete  modeling  of a PPN-W formation   requires   simulating an RGB or AGB in RLOF with a companion accretor, and  allowing the field to amplify within the  accretion flow. A self-consistent magnetically mediated outflow should grow,  along with any circumbinary molecular torus that might aid in collimation.  
For self-consistent  PPN-C formation,  a  simulation would instead require the  full CEE after the RLOF phase,  starting with a stellar seed magnetic field in the giant and computing how the combination of CE, inspiral, field amplification, outflow, and collimation  subsequently proceed.

There has not yet been a fully  self-consistent simulation of a PPN-W by the above standard, and so here I focus on PPN-C.  Indeed, \cite{Ondratschek+2021} have broken new ground for the study of  PPN-C with a  full MHD  CEE simulation that shows organic growth of  a jet mediating magnetic field and collimated outflow production in 3-D.  They use the  AREPO code and include  a  prescription for recombination energy, which is important as it causes the simulated envelope to become unbound during the jet evolution. They start with a weak dynamically insignificant seed magnetic dipole field in the AGB star of mass $0.97\ {\rm M}_\odot$ and
include a companion of $0.243\ {\rm M}_\odot$ as fiducial run.  This mass ratio is then $q=0.25$, but they  also carried out runs for $q=0.5$ and $q=0.75$  that produced qualitatively similar results.
They track the inspiral to  $\sim 4000$
days. 


The  magnetic field is amplified some 15 orders of magnitude  on the time scale of the simulations. Comparing runs with  and without a magnetic  field, they find that the field makes little difference to unbinding  after 1000 days  (also true of semi-analytical dynamos, \citeauthor{Nordhaus+07}\citeyear{Nordhaus+07}), which is expected: as discussed above, the magnetic field is not an extra source of energy but draws its energy from the sources already there, the  orbital energy in this case.  The field amplification arises from some combination of shear, and turbulence sourced by  some combination of MRI,  and Kelvin-Helmholtz instability and the emergent jet launches with cross sectional diameter of the circumbinary disk engine. The  exact analytical modeling of the  system has yet to be carried out,  but this  clearly demonstrates a type of PPN-C outflow,  as the jet emanates from a circumbinary region which is a factor of $\sim 8$ times smaller than the initial AGB stellar radius. 
The most important lesson that this simulation highlights, is that   the magnetic field facilitates  formation of a magnetically driven and strongly collimated jet along a narrow axial channel.  The collimated jet is absent without the field.   

The simulation by \cite{Ondratschek+2021} is a substantial step toward high fidelity global simulations of PPN-C.  Under the hood, there are a number of issues that warrant further work and discussion. Convection in the AGB star has not been included. The treatment of recombination is approximate and non-local,  and the extent to how this interfaces with convection is  relevant in the broader CE context as discussed earlier. Convection will also add  a disordered component  to the initial seed dipole field and so the extent to which it influences the overall magnetic launch and collimation is also important. In the presence of convection, even the seed field may require a dynamo in the star since exponential generation is needed to compete with turbulent diffusion.
There is also the question as to the relative role of  a magnetic tower or a magneto-centrifugal launch, a distinction discussed further in the next section.

\subsection{Magnetic Tower, Magneto-centrifugal Launch, or Magnetic Bomb?}

Magnetically mediated launches that depend on the presence of large-scale magnetic fields are not all the same. There are essentially three types as described below. Although all  share the fact that  a gradient in toroidal field magnetic pressure helps to propel material and the hoop stress helps to collimate, they also differ in key aspects. They may not be mutually exclusive in a given source. For example,   a  magnetic tower may be embedded within a broader magneto-centrifugual launch, however, we do not yet know which mechanism dominates in any given source.


The magnetic tower (MT)  \citep[e.g.][]{lynden-bell1996,lbl2003,Uzdensky+2006,Gan+2017}
can be initiated from   magnetic field loops anchored between footpoints in relative differential rotation, for example loops that link a stellar core to a surrounding  torus.  The footpoint separation may be of comparable  scale  to the radius at which they are anchored, but  both footpoints are contained in the engine itself.
The differential rotation winds up the field,  creating a toroidal component that establishes a magnetic pressure force which  
pushes material upward.  The hoop stress  collimates the pressure of this rising tower, but only when the ambient medium is surrounded by a balancing ambient pressure. The magnetic field  is parallel  to the axial flow on the jet axis, and becomes increasingly toroidal away from the axis toward the jet boundary.  
The outflows can remain marginally magnetically dominated inside the jet out to observable scales within the tower and maybe out to arbitrarily large-scales.  Importantly, both signs of magnetic flux are contained within the structure that would appear as the jet tower because both inner and outer footpoints of the original loops that form  tower are within the jet.  Their separation defines the diameter at the base of the tower.  The jet contains zero vertical net magnetic flux. 


The magneto-centrifugal launch (MCL) 
\citep{blandford1982,Pelletier+1992}
is a  more  widely invoked class of models for  which the starting point is a large-scale open magnetic field  with only one anchoring 
foot-point sign at the engine base and the other foot-point essentially at infinity.  Plasma loaded onto quasi-rigid field lines at the base is centrifugally flung along these lines as angular momentum of the base is transferred via the field lines to the plasma.  At the Alfv\'en radius, where the flow energy density becomes comparable to that of the field, the torodial field magnitude is comparable to that of the poloidal field and supplies some outward pressure and collimation.  
On larger scales farther from the engine, say $\gtrsim 100 R_{in}$, the  flow kinetic energy marginally dominates field energy within the jet,  the opposite of the magnetic tower. 
Importantly, for the MCL,  only one sign of poloidal magnetic flux resides within the jet, also in contrast to magnetic tower. 
For a source of magnetic field  produced in a PPN accretion disk, MCL outflow scalings have  been applied to PPN
\citep{Blackman+2001}. 
For the MCL like the MT,  ambient collimation of the inner magnetic structures by some ambient flow or pressure is also  needed and all simulations demonstrating collimation and steady jets have pressure equilibrium at the jet  boundary.

One more  magnetically mediated outflow model is a magnetic bomb (MB)  \citep{Matt+2006}.  Here a wound magnetic field acts like a capacitor, suddenly releasing its energy in outflows.   The starting point is an evolved star for which differential  rotation has been established between the collapsed degenerate core, and the expanded envelope.  The initial field anchored at the core is open, as per the MCL.  However,  this field can be  weak initially,  and there is no need to load mass from the core onto the field lines as  the field is already mass loaded. Differential rotation between the envelope and core winds up  the toroidal field. After the field  reaches some threshold, 
it rapidly drives  polar outflows from vertical magnetic pressure gradients, and equatorial outflows as the wound field in each hemisphere also squeezes flow outward from the equator.  Like the MT,   differential rotation is important from the start,  but the primary differential rotation for the MB is vertical not radial.  
The MB for an initial  dipole field would have just one sign of magnetic flux in the outflow, similar to the MCL and distinct from the MT.

\subsection{Measured Magnetic Fields}

Given the importance of magnetic fields, some further comments on what has been  measured warrants  mention.
The measurement of magnetic fields in W43A
\citep{Vlemmings+2006,Amiri+2010} and the evidence for toroidal fields in the Calibash from polarization \citep{Sabin+2015} were mentioned above.   Fields have also been  measured in other AGB stars via masers \citep{Herpin+2006},  masers in the OH shell of  NML Cygni \cite{Etoka+2004}, and in Miras 
\citep{Kemball+1997,Kemball+2009}.
Synchrotron emission in the post AGB star IRAS 15445-5449 has  been measured  at 7000 au, indicating a mG field which could collimate the jet  \citep{Perez-Sanchez+2013}.
\cite{Leone+2014} found no evidence for  $>$ kG fields in the central stars of PN, but a field of $B\sim 655$G in NGC4361. 

However, an important  point  is that none of the above measurements  probe the field on the jet  ``launch" scales ($< 0.01$ au). Instead, they are measurements  of the jet ``propagation” ($> 1$ au) scales. That fields are dynamically important on propagation  scales is important, but these do not directly constrain the jet formation at its engine. 

Also, distinguishing the MT from MCL cannot be done with polarization measurements because they are ambiguous to 180 degrees in field orientation.
Measurements  that constrain whether the sign of the flux is uniform across the jet in a given hemisphere (indicating MCL or MB)   or has two reversals across the jet (indicating MT) are needed.

\section{Persistent  challenges and distinguishing PPN-C from PPN-W}

While PPN and CEE are likely associated, one challenge
is to identify direct signatures of a CEE event 
\citep{Khouri+2021} and possibly to associate the two
more directly.  There remain some challenges that come with comparing theory, simulation and observation both in terms of the limits of numerical simulations and in predicting  distinguishing features of  PPN-C and PPN-W.

\subsection{Convection, magnetic fields, and the limits of numerical simulations}
In addition to potentially delocalising the deposition of recombination energy  discussed earlier, convection also leads to turbulent diffusion of large-scale magnetic flux. In turbulent astrophysical systems,   large-scale flux is almost never frozen. (Think of the Sun,  where the large-scale field reverses every 11 years. This would be  impossible if flux were frozen.) 
This raises the question of the initial field in the giant star used for simulations without convection \cite[e.g.][]{Ondratschek+2021}.  In reality, 
the stellar field would not be  exclusively  ordered  but may even be dominated by a random component. The total  field would need to be sustained by large and small scale dynamos to overcome the exponential decay from turbulent diffusion. How   various dynamos  conspire in these engines to produce the dynamically significant large-scale fields, and how the turbulence affects the level of collimation of the jet remains to be studied.

 A second open issue is that large scale transport rather than local isotropic  turbulence 
may  dominate  in disks, whether or not the magneto-rotational instability (MRI) is the dominant instability  \citep{BNreview}.  The fraction of small scale, mesoscale, or large-scale  transport is not well constrained.   The scale of transport  determines where  energy is dissipated,  which in turn determines  the observed spectral signatures of accreting systems. Dissipation that occurs deep within an optically thick disk will produce thermal emission but  dissipation that  occurs in a corona or jet can be non-thermal.  The fraction of thermal versus non-thermal emission can thus be used  as a proxy for the scale of angular momentum transport \citep{blackmanpessah}.

Another pervasive issue, is how sensitive  the phenomenological
output from simulations is as a function of  initial and boundary conditions.   Numerical simulations are useful to study small pieces of a physical system at high resolution or ``kitchen sink" approaches at low resolution.  But for kitchen sink simulations, the question of convergence is  substantial.  Dynamos  and  accretion disks  are  examples   for which  intermediate fidelity simulations  can cause confusion. Suppose, for example, that analytic theory were to predict that in the asymptotic limit of large magnetic Reynolds number,  a particular dynamo magnetic growth rate  is independent of magnetic Reynolds number.  Further suppose that  the  real astrophysical system of interest has a magnetic Reynolds number much larger than one could ever hope to simulate. If a simulation then exhibits a dependence on magnetic Reynolds number, it is not easy to determine whether the theory is wrong or whether the simulation is not in a sufficiently asymptotic regime to be realistic. In the case of the solar dynamo, intermediate fidelity simulations have indeed sometimes produced results that disagree with first generation theory,  whilst higher fidelity simulations have shown more consistency with basic theory.  In short, basic theory and computational simulations represent distinct approaches that are valuable both independently and   in combination.

\subsection{Identifying distinct features of PPN-C and PPN-W }

 The classification of PPN-W versus PPN-C, respectively, delineates whether the influential binary companion is  inside or outside the outer radius of the initial giant star. Which observational consequences manifest  from this distinction? In addition to subtle differences that may require  simulations to  identify, there are some  conspicuous distinctions.

PPN-W would show the time variability of a larger binary orbit than PPN-C. Moreover, for PPN-W the jet has a  diameter at launch that is much smaller than the orbital separation, so one might expect reflection symmetry to be common as the jet moves around the orbit.  In contrast, for PPN-C, the jet cross section is closer to the size of the binary separation itself and less  orbital motion of the jet is expected.  There is also likely to be more surrounding torus mass for PPN-C than for PPN-W since the former happens inside the circumstellar envelope.  This would suggest that more collimation is likely for PPN-C than PPN-W.  A PPN-C jet may thus  be  narrower, possibly with  point symmetry rather than reflection symmetry, if the overall jet precesses. 

There may also be  statistical population differences in duration between PPN-C and PPN-W as they are determined by different accretion processes. 
PPN-W durations would be determined by how long RLOF accretion  occurs before the companion plunges into CEE. 
Predictions for this duration are not yet clear
\citep{Macleod+18b}. 
PPN-C would be determined by the time scale for  circumbinary material to accrete after CEE but before enough envelope is lost to reduce the mass supply, or the time scale for accretion from a merger to run out of mass. 

There may also be some influence of WD nuclear burning
that is more common for PPN-C  than PPN-W because all PPN-C would have at least 1 WD within its engine.   Core X-rays and other signatures of dwarf novae could be more prevalent for PPN-C.

Another distinction for single companion interactions is that although RLOF fueled PPN-W can produce a surrounding torus that precedes a jet, only for PPN-W can there also be a torus that  follows such a jet. This is because the PPN-W happens before CEE, if CEE is to happen, whereas  PPN-C would always occur after CEE. This distinction could lead to 
 statistical differences between the timing of jet and tori in the two populations.  As emphasized earlier,  even magnetized jets require  an ambient wind or torus for stable collimation, so some minimum ambient material would be common to both PPN-C and PPN-W. Multiple companion interactions could complicate this trend if, for example,  a first companion produces a visible jet after inspiral but does not  eject the envelope, and a second insipiral and jet follow.

Finally there may be compositional differences. Since material supplying PPN-W jets comes from  farther in the envelope,  it may be   less C-rich.  The jet composition will be dependent  on the jet material source, and this also depends on how effectively or ineffectively convective mixing homogenizes the composition.

\section{Summary}

PPN are more kinematically demanding than PN,  and thus 
place tighter constraints on their mutual origin mechanisms  if  PN are  a time evolved state of PPN.   Kinematic constraints for PPN demand close binary interaction, at least as close as RLOF from the primary onto a secondary main-sequence star. 

To produce PPN jets, binaries and magnetic fields likely act in symbiosis,  with the free energy in orbital motion and accretion  used to generate magnetic fields that drive and collimate outflows.  The presence of circumstellar tori can in turn collimate and stabilize the magnetic structures, which is likely required to explain observed PPN jets.
The classification of PPN jets as either PPN-W or PPN-C can be used to respectively distinguish those for which the binary separation in the engine is wider than, or less than the primary  giant envelope radius. Jets produced by RLOF of the giant onto the secondary are examples of PPN-W, and jets produced after the companion plunges into CEE by a circumbinary disk or merger after the envelope largely unbinds would be PPN-C  jets.  Examples of both classes
were discussed.

There has been significant progress over the past several decades in putting all of the pieces together. This is  culminating  in increasingly high fidelity simulations.   
Self-consistently generated magnetic fields, and the associated magnetically mediated jets 
are now seen to  emerge organically in   3-D CE PPN-C simulations. There is open opportunity to  achieve the equivalent for PPN-W. 

Probing the physics  of these simulations and their limitations remains an active effort. Convection is a fundamental feature of observed giants that is absent from most simulations, and  presents an important frontier. The extent to which the observed phenomenology depends on initial binary parameters and boundary conditions needs to be explored.  Understanding the different classes of PPN that can  be produced and their observational signatures will benefit from further collaborations between observers and theorists to converge on specific predictions that can  distinguish different mechanisms.

\bibliographystyle{apj}

\section*{Acknowledgements}
Thanks to L. Chamandy for useful discussions and comments, and also to B. Balick,  O. De Marco, A. Frank, J. Kastner, J. Kluska,  J. Nordhaus, R. Sahai,  N. Soker,  E. Wilson, and A. Zou for pertinent discussions. 
Thanks to L. Decin for meticulous editing and  comments.
Support  from  US Department of Energy grants 
DE-SC0020432 and DE-SC0020434,
and  US National Science Foundation grants 
AST-1813298, and  PHY-2020249  are acknowledged.

\bibliography{IAU366refs.bib,general.bib}
\end{document}